\documentclass[11pt]{article}
\usepackage[utf8]{inputenc}
\usepackage{graphicx,amsmath,amsfonts,amssymb,fullpage,xcolor,booktabs}

\usepackage[T1]{fontenc}

\usepackage[numbers, compress, super]{natbib}

\RequirePackage{hyperref}
\hypersetup{colorlinks=true, citecolor=black, linkcolor=black, urlcolor=black}

\usepackage[most]{tcolorbox}
\newtcolorbox[
  auto counter,
]{kbox}[2][]{
  enhanced, 
  coltitle=black, 
  fonttitle=\bfseries,
  attach title to upper={\ },
  title=Box \thetcbcounter: #2, #1
}

\title{\textbf{Key-value memory in the brain}}
\author{Samuel J. Gershman$^{1,2,3,\ast}$, Ila Fiete$^{4}$, and Kazuki Irie$^{1}$ \\
$^1$Department of Psychology\\
$^2$Center for Brain Science\\
$^3$Kempner Institute for the Study of Natural and Artificial Intelligence, \\Harvard University, Cambridge, MA, USA\\
$^4$McGovern Institute for Brain Research and Department of Brain and Cognitive Sciences, \\Massachusetts Institute of Technology\\
$^\ast$Corresponding author: gershman@fas.harvard.edu}

\begin{document}

\maketitle

\begin{abstract}
    Classical models of memory in psychology and neuroscience rely on similarity-based retrieval of stored patterns, where similarity is a function of retrieval cues and the stored patterns. While parsimonious, these models do not allow distinct representations for storage and retrieval, despite their distinct computational demands. Key-value memory systems, in contrast, distinguish representations used for storage (values) and those used for retrieval (keys). This allows key-value memory systems to optimize simultaneously for fidelity in storage and discriminability in retrieval. We review the computational foundations of key-value memory, its role in modern machine learning systems, related ideas from psychology and neuroscience, applications to a number of empirical puzzles, and possible biological implementations.
\end{abstract}

\section{Introduction}

Despite the apparent fragility of memory, there is no decisive evidence that information, once stored, is ever permanently lost. The storage capacity of the brain of course must be finite, but that does not seem to be the principal limiting factor on memory performance. Rather, it is the retrieval process that fundamentally limits performance: the relevant information may be there, but cannot always be found \citep{tulving74,crowder76,lewis1979psychobiology,miller21}. Some of the evidence supporting this view will be summarized below.

A retrieval-oriented view of memory performance places the primary explanatory burden on how memories are \emph{addressed} (i.e., how the retrieval system keeps track of storage locations), and how they are \emph{queried} (i.e., how the retrieval system maps sensory cues to addresses). There is a long history of theorizing about these concepts in cognitive psychology and neuroscience \citep{kahana12,chaudhuri16}. Recently, the field of machine learning has developed its own analysis of these concepts, which form the basis of high-performing systems like transformers and fast weight programmers \citep{vaswani17,schmidhuber1992learning}. 
There is increasing recognition that a significant aspect of intelligence (in both natural and artificial systems) is effective information retrieval \citep{graves16,gershman17,geva21,irie22,allen-zhu24}.

Our goal is to connect the dots between conceptualizations of memory retrieval in psychology, neuroscience, and machine learning. Central to this effort is the concept of \emph{key-value memory}, which we formalize below. The basic idea is that inputs (memoranda) are transformed into two distinct representations---keys and values---which are both stored in memory. The keys represent memory addresses, while the values store memory content. Memories are accessed by first matching a query to each key, and then retrieving a combination of values weighted by their corresponding matches. Importantly, the mappings from inputs into keys and values can be optimized separately, allowing the system to distinguish between information useful for finding memories (stored in the keys) and information useful for answering the query (stored in the values). This distinction is familiar in human-designed information retrieval systems. For example, books often have alphabetically organized indices, which are helpful for finding particular subjects. The indices do not contain any information about the meaning of the subjects themselves; this information is stored in the book's text, retrieved by going to the page number associated with the index.

We will argue that memory in the brain follows similar principles. In particular, we posit a division of labor between a key storage system in the medial temporal lobe and a value storage system in the neocortex. As reviewed below, closely related ideas have already been proposed in psychology and neuroscience. By connecting these ideas to key-value memories, we can begin to understand what makes memory in the brain computationally powerful. To illustrate these points, we present simulations that recapitulate a number of empirical phenomena. Implications for the convergence of natural and artificial intelligence are discussed in the concluding section.

\section{Computational foundations of key-value memory}
\label{sec:key-value}

In this section, we introduce the technical ideas underlying key-value memory, exposing the multitude of ways in which this idea has been conceptualized. We place key-value memory within a modern machine learning framework by discussing how the key and value representations can be learned, comparing an end-to-end learning approach with fixed (or partially fixed) ``scaffolds'' for key representations.

\subsection{From correlations to kernels}

\begin{figure}
    \begin{center}
    \includegraphics[width=1\textwidth, scale=1]{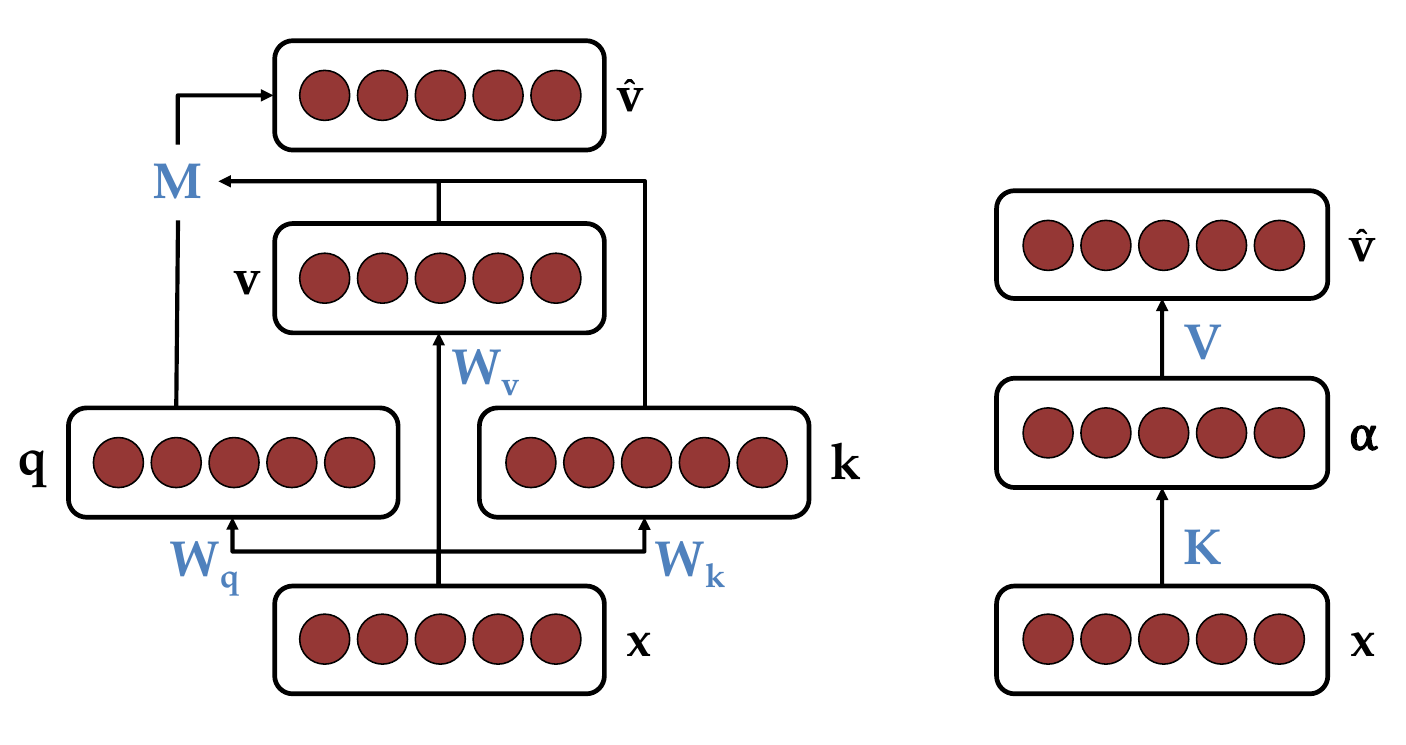}
    \end{center}
    \caption{\textbf{Two architectures for key-value memory}. Black symbols denote vectors and blue symbols denote matrices. (Left) Input $\mathbf{x}$ is mapped to key ($\mathbf{k}$), query ($\mathbf{q}$), and value ($\mathbf{v}$) vectors. During memory writing, the weight matrix $\mathbf{M}$ is updated using Hebbian learning between the key and value vectors. During reading, the query is projected onto $\mathbf{M}$ to produce a retrieved value $\hat{\mathbf{v}}$. (Right) The input vector is mapped to a hidden layer $\boldsymbol \alpha$, which is then mapped to an output layer $\hat{\mathbf{v}}$. The input-to-hidden weights correspond to the stored keys; the hidden-to-output weights correspond to the stored values.}
    \label{fig:schematic}
\end{figure}

One of the earliest formalizations of a key-value memory was Kohonen's correlation matrix memory model \citep{kohonen72}, which was subsequently used by Pike \citep{pike84} to explain a range of human memory phenomena. Here we slightly change the notation and terminology to bring this model into correspondence with more recent formalizations (Figure \ref{fig:schematic}, left). Each input, indexed by $n$, consists of a key vector $\mathbf{k}_n$ and a value vector $\mathbf{v}_n$ (which we take to be row vectors). Intuitively, the keys encode information about memory addresses (hence we will refer to the set of possible keys as the \emph{address space}), while the values encode information about memory content. These two representations are linked in memory by an ``associator'' matrix $\mathbf{M}$, which is initialized at 0 and incremented by the outer product of the key and value vectors after each input is presented:
\begin{align}
    \Delta \mathbf{M} \propto \mathbf{k}_n^\top \mathbf{v}_n.
    \label{eq:hebb}
\end{align}
It is easy to see that this is just simple Hebbian learning between key and value units encoding the vector elements. In neurobiology, the standard interpretation is that $M_{ij}$ is encoded by the synaptic strength between neurons representing value element $j$ and key element $i$. The synaptic strength is increased when the two neurons fire coincidentally \citep{caporale08}. Needless to say, learning in the brain is more complicated than this, but for present purposes we will take for granted the biological plausibility of Eq.~\ref{eq:hebb} (see also \citep{limbacher20} for a related, but more complex, learning rule).

The correlation matrix memory is \emph{heteroassociative} because $\mathbf{M}$ stores information about the relationship between two different kinds of objects or object properties (see also \citep{steinbuchP63} for one of the earliest such models). If we impose the constraint that keys and values are the same, we get an \emph{autoassociative} memory \citep{willshaw1969non,anderson70,amari1972,nakano72}. This idea is now most closely associated with the work of Hopfield \citep{hopfield82}, and such models (when the inputs are binary) are typically referred to as \emph{Hopfield networks}.

The correlation matrix memory is queried by taking the inner product between the associator matrix and a query vector $\mathbf{q}$ (with the same dimensionality as the key vectors):
\begin{align}
    \hat{\mathbf{v}} = \mathbf{q} \mathbf{M}.
    \label{eq:retrieval}
\end{align}
In neural network terms, this is equivalent to activating pattern $\mathbf{q}$ on the key units, which produces an activity pattern in downstream value units based on the learned synaptic strengths ($\mathbf{M}$). The retrieved value vector $\hat{\mathbf{v}}$ combines stored values associated with keys that are similar to the query vector. To see this, it is helpful to rewrite Eq.~\ref{eq:retrieval} in its ``dual'' form \citep{ba2016using,katharopoulos20}:
\begin{align}
    \hat{\mathbf{v}} \propto \sum_{n=1}^N \alpha_n \mathbf{v}_n,
\end{align}
where $\{\alpha_n \}$ is a set of ``attention weights'' computed by $\alpha = \sigma(S(\mathbf{K},\mathbf{q})) =  \mathbf{q} \mathbf{K}^\top$,
and $\mathbf{K}$ is the matrix consisting of all $N$ key vectors (i.e., its rows are $\mathbf{k}_n$). The function $S(\cdot,\cdot)$ is a \emph{similarity kernel} (linear in this case), expressing the match between keys and queries, and $\sigma(\cdot)$ is a \emph{separation operator} (the identity function in this case), pushing apart similar memories. The retrieved value vector is thus a weighted combination of stored values, where the attention weights correspond to the similarity between the query and each key, mapped through a separation operator.

The dual form is useful because it shows how the model can be straightforwardly generalized by considering other similarity kernels and separation operators. Any positive semidefinite kernel can be expressed as the inner product in some vector space \citep{scholkopf02}, which means that alternative similarity functions can be constructed by mapping the keys and queries through a feature transform $\phi(\cdot)$ applied to each row:
\begin{align}
    S(\mathbf{K},\mathbf{q}) = \phi(\mathbf{q}) \phi(\mathbf{K})^\top.
    \label{eq:raw_score}
\end{align}
It is also possible to construct kernels with infinite-dimensional vector spaces by working directly with an explicitly defined similarity kernel, such as the widely used radial basis function kernel, which Tsai and colleagues \citep{tsai19} found to produce the best results.

The generalized correlation matrix memory can implement the celebrated self-attention mechanism in transformers \citep{vaswani17}:
\begin{align}
    &S(\mathbf{K},\mathbf{q}) = \frac{\mathbf{q} \mathbf{K}^\top}{\sqrt{D}} \label{eq:scaled_dot_prod} \\
    &\sigma(\tilde{\alpha}) = \frac{\exp(\tilde{\alpha})}{\sum_n \exp(\tilde{\alpha}_n)}, \label{eq:softmax}
\end{align}
where $D$ is the dimensionality of the key/value vectors, and the softmax function is used as the separation operator. Here we have used $\tilde{\alpha}$ to denote the attention weights prior to the application of the separation operator. When $\sigma(\cdot)$ is set to the identity function, we recover linearized attention \citep{katharopoulos20}, which can be expressed as a special form of recurrent neural network known as a \emph{fast weight programmer} \citep{schmidhuber1992learning,schlag21}. Linearized attention has a potential computational advantage over softmax attention, because its recurrent form allows for linear-complexity sequence processing, while giving up the parallel computation property of the attention form (whose complexity, however, is quadratic in the number of sequence elements). Several studies \citep{choromanski21,peng2021random} have shown how to rigorously approximate softmax attention using random features.

As shown by Millidge and colleagues \citep{millidge22}, a variety of classical memory models can also be derived by different choices of similarity kernel and separation function. For example, sparse distributed memory \citep{kanerva88} can be obtained by setting $\sigma(\cdot)$ to a threshold function, and dense associative memory \citep{krotov16} can be obtained by setting $\sigma(\cdot)$ to a rectified polynomial.

In the noiseless setting, the ideal separation function is the max operator, since this will always return the stored value associated with the matching key \citep{millidge22}. However, this separation function is not robust to noise---small perturbations can cause large retrieval errors. Thus, the design of a memory system needs to balance separability and robustness. In addition, the memory may be used for generalization to new inputs \citep{vaswani17,ramsauer21,bricken23}, in which case perfect matching is not the goal. For example, key-value memories have been used extensively for question-answering tasks \citep{sukhbaatar15,miller16}, where $\mathbf{q}$ represents a question and $\mathbf{v}$ represents an answer. The system stores a set of question-answer pairs and tries to use this knowledge base to answer novel questions.

\subsection{Representational structure}

So far, our setup has assumed a memory system provided with keys, values, and queries---where do these representations come from? In modern machine learning, they are derived as mappings from input vectors, $\{\mathbf{x}_n\}$. A typical assumption in transformers and fast weight programming is to parametrize the mappings with a set of linear functions:
\begin{align}
    &\mathbf{k}_n = \mathbf{x}_n \mathbf{W}_k \\
    &\mathbf{v}_n = \mathbf{x}_n \mathbf{W}_v \\
    &\mathbf{q}_n = \mathbf{x}_n \mathbf{W}_q.
\end{align}
where $\{\mathbf{W}_k, \mathbf{W}_v, \mathbf{W}_q \}$ are learned weight matrices. The input vectors may themselves be learned embeddings of raw input data, harnessing the power of deep learning systems trained end-to-end.

In some systems, the key and query mappings are assumed to be fixed, either randomly or based on some regular structure. In these systems, the address space is conceived as a kind of ``scaffold'' for the indexing of information content. In the sparse distributed memory model \citep{kanerva88}, the scaffold is random, so that similar keys do not systematically index similar values. This mimics (approximately) the organization of random access memory in digital computers. A familiar example of a structured scaffold is the alphabetical index found at the end of books. In the Hopfield network \citep{hopfield82} and related autoassociative memory models, the scaffold is identical to the value space. This property endows the models with \emph{content addressability} \citep{kohonen80}: memory addresses are accessed by matching values directly to queries. Most psychological and neural models of memory share this property, though they implement it in different ways.

Some models assume that the scaffold is randomly constructed in such a way that it carries a structural imprint. For example, if inputs are assigned to random addresses, but these addresses change slowly over time, then inputs experienced nearby in time will encoded as more ``similar'' in the address space. Temporal autocorrelation, in the absence of additional structure, is able to account for many aspects of human \citep{landauer75} and animal \citep{estes55} memory.

Below, we will discuss neurobiologically inspired architectures that implement a fixed scaffold, which can even outperform a learned mapping. We will also show that learning the key and query mappings allows us to explain data on repulsion effects in memory and hippocampal representations. These results pose a number of questions for future research: Does the brain use a combination of fixed and learned mappings? Is such a combination useful for machine learning applications?

\subsection{The ubiquity of key-value memory}
\label{sec:ubiquity}

The work reviewed above focused on explicit constructions of key-value memory. It turns out that other models are sometimes implicitly equivalent. Irie and colleagues \citep{irie22} showed that linear layers trained by gradient descent (ubiquitous in many machine learning models) can also be expressed as key-value memories. Here we briefly review this theorem.

Let $\mathbf{x}_n$ be the input vector at time $n$, as in the previous section. The output of a linear layer is
$\mathbf{y}_n = \mathbf{x}_n \mathbf{W}$, where $\mathbf{W}$ is as weight matrix. The weight matrix is trained by gradient descent, yielding the following after $N$ timesteps:
\begin{align}
        \mathbf{W} = \mathbf{W}_0 + \sum_{n=1}^N \mathbf{x}_n^\top \mathbf{e}_n,
\end{align}
where $\mathbf{W}_0$ is the initial weight matrix, and $\mathbf{e}_n = -\eta_n (\nabla_{\mathbf{y}} \mathcal{L})_n$ is the error signal with learning rate $\eta_n$ and loss function $\mathcal{L}$. Generalizing a classic result on kernel machines \citep{aizerman64}, Irie and colleagues \citep{irie22} showed that this construction is equivalent (in the sense of producing the same output for a given input) to the following linear key-value memory:
\begin{align}
    \mathbf{y} = \mathbf{x}\mathbf{W}_0 + \sum_{n=1}^N \alpha_n \mathbf{v}_n,
\end{align}
where (using our earlier notation) $\alpha = \mathbf{q} \mathbf{K}^\top$ 
with $\mathbf{v}_n = \mathbf{e}_n$, $\mathbf{k}_n = \mathbf{x}_n$, and $\mathbf{q} = \mathbf{x}$, given an arbitrary input $\mathbf{x}$. Thus, linear layers effectively memorize their experienced error patterns, computing their outputs as linear functions of these memories. This interpretation is intriguing in light of the fact that errors are particularly salient in human memory \citep{green56,hirshman89,sakamoto04,rouhani18,bein21}.

The linear layers retain information about all training inputs---they never ``forget'' \citep{irie22}. However, retrieval access may be (transiently) lost. We discuss evidence for this idea from psychology and neuroscience below.

\section{Neurobiological substrates}

While key-value memories are loosely inspired by the brain, it remains an open question how to implement them in a biologically plausible manner. Eq.~\ref{eq:hebb} is a (somewhat) biologically plausible rule for learning associations between keys and values, but we still need rules for storing the keys and values themselves. In the section on representation learning, we described the widely used approach of modeling keys, queries, and values as linear mappings from input vectors (which themselves could be learned). This implies additional learning rules for the linear mappings. One could posit that they are learned by backpropagating the errors from whatever objective function is being optimized for a downstream task \citep{limbacher20}. This would require a biologically plausible approximation of the backpropagation algorithm \citep{lillicrap20}.

Kozachkov and colleagues \citep{kozachkov23} have proposed an architecture based on the ``tripartite synapse'' consisting of pre-synaptic and post-synaptic neurons modulated by an astrocyte (a type of glial cell). In particular, they posited that this motif applies to the hidden-to-output synapses in the three-layer network implementation of key-value memory. The activation of each astrocyte process is modeled as a linear function of the hidden unit activations. They showed how the astrocyte processes collectively compute the similarity function $S(\mathbf{K},\mathbf{q})$, which then multiplicatively modulate the hidden-to-output weights so that the network as a whole implements the transformer self-attention described earlier.

Alternatively, Tyulmankov and colleagues \citep{tyulmankov21} have proposed a non-Hebbian learning rule for key learning. They view the key-value memory as a three-layer neural network, where the input $\mathbf{x}$ (first layer) is transformed into a pattern of attention $\boldsymbol \alpha$ (hidden layer), which is finally transformed into retrieved values $\hat{\mathbf{v}}$ (output layer); see the right panel of Figure \ref{fig:schematic}. Each hidden layer unit represents a ``slot'' to which a single input (or possibly multiple inputs) gets assigned. In this view, the key matrix $\mathbf{K}$ corresponds to the input-hidden synaptic strengths, and the value matrix $\mathbf{V}$ (where row $n$ corresponds to value vector $\mathbf{v}_n$) corresponds to the hidden-output synaptic strengths. The proposed learning rule for the synapse connecting input unit $j$ to hidden unit $i$ is (simplifying slightly, and dropping time indices):
\begin{align}
    \Delta K_{ij} \propto \mu \gamma_i (x_j - K_{ij}),
\end{align}
where $\mu \in \{0,1\}$ is a global third factor (possibly a neuromodulator like acetylcholine or dopamine) and $\gamma_i \in \{0,1\}$ is a local third factor (possibly a dendritic spike). The factors are binary to promote sparsity in the hidden layer. The local third factor tags the least-recently-used hidden unit as eligible for plasticity. The authors point out that the key learning rule resembles behavioral time scale plasticity \citep{bittner17}, which has been observed in the hippocampus. The possibility that key learning occurs in the hippocampus will be considered in detail below.

For value learning, Tzyulmankov and colleagues clamp the output layer to the target value $\mathbf{v}$ and then update the synapse connecting hidden unit $i$ to output unit $m$ according to:
\begin{align}
    \Delta V_{mi} \propto \mu \gamma_i \alpha_i (v_m - V_{mi}).
\end{align}
This is a Hebbian rule, because it depends on the co-activation of hidden and output units.  It could potentially describe modification of the output projections from the hippocampal area CA1 to entorhinal cortex.

Using a similar architecture (but with the important addition of recurrence), Whittington and colleagues \citep{whittington22} have shown that the Tolman-Eichenbaum Machine \citep{whittington20}, a model of the entorhinal-hippocampal system, implements a form of key-value memory. The hippocampus, according to this model, stores conjunctive representations of sensory inputs from lateral entorhinal cortex and a ``structural code'' provided by cells in medial entorhinal cortex, which are updated recurrently. Conjunctive representations in the hippocampus function as both the keys and the values (i.e., an autoassociative memory). In a 2D open field, the model can reproduce hippocampal ``place cells'' (firing fields localized to a particular spatial location), while the structural code generates grid cell-like responses (grid cells fire with hexagonal periodicity as an animal moves through space \citep{hafting05}). In environments of different dimensions, topologies, and geometries, the structural code need not be grid-like. Whittington and colleagues note that the structural code in their model functions as a form of ``position encoding''---widely used in transformers and other sequence models---where the encodings (i.e., addresses) are adapted to the structure of space. This is a critical aspect of the model: the structure-sensitive properties of their modeled hippocampal and entorhinal cells would not have been obtained using the fixed positional encodings typically used in regular transformers. Related ideas have proven successful in the machine learning literature \citep{liu20}.

\begin{kbox}[boxrule=0pt, parbox=false, opacityframe=0, label=box:hashmesh]{Memory with error-correcting key-query matching}
\\ \newline
Vector-HaSH \citep{chandra25} and MESH \citep{sharma22} are tripartite networks consisting of an input (sensory) layer $s$, a densely connected layer $d$, and a layer consisting of a modular set of fixed recurrent attractor networks $a$. These models dissociate content storage from query-matching and directly translate to a key-query-value formulation. Define the tripartite network weights to be
${\bf W}_{sd}, {\bf W}_{da}, {\bf W}_{ad}, {\bf W}_{ds}$, and the attractor, dense layer, and sensory states to be $\{{\bf a}\}, \{{\bf d}\}, \{{\bf s}\}$, respectively. Given some sensory input ${\bf s}$, the network retrieves the nearest relevant stored state $\hat{\bf s}$ by performing the following operations (treating the dense layer responses as linear, though in practice they are nonlinear): 
\[
\hat{{\bf s}} = \phi({\bf s}{\bf W}_{sd}{\bf W}_{da}, \{{\bf a}\}) {\bf W}_{ad}{\bf W}_{ds}. 
\]
Here $\phi(.,.)$ represents an efficient nearest-neighbor (NN) operation performed by the scaffold (combined dense and attractor layers), to select the closest attractor state for the given input. The readout weights $W_{ad}, W_{ds}$ reconstruct the sensory memory from the selected attractor state. This model maps to the key-value system of Figure \ref{fig:schematic}B: the output of the attractor layer is a (denoised) query 
$${\bf q}_n = \phi({\bf s}{\bf W}_{sd}{\bf W}_{da}, \{{\bf a}\}) \equiv {\bf x}_n $$ 
and the key and value matrices are given by the attractor-to-dense and dense-to-sensory weights, repsectively, $\mathbf{K} = {\bf W}_{ad}^\top, \ ~ \mathbf{V} = {\bf W}_{ds}.$ Thus, the scaffold computes a similarity $\sigma(S({\bf K},{\bf q}))$ and the network produces the output 
$$\hat{{{\bf v}}}_n = \hat{{\bf s}}_n = \alpha_n {\bf W}_{ds} = \phi({\bf s}_n{\bf W}_{sd}{\bf W}_{da}, \{{\bf a}\}){\bf W}_{ad}{\bf W}_{ds} = \sigma(S({\bf K},{\bf q}_n)){\bf V}$$
where the attention weights (Fig. \ref{fig:schematic}B's hidden states) are $\alpha_n = {\bf x}_n {\bf W}_{ad}$.

The value weights (from the dense to the sensory layer) and part of the query weights (from the sensory to the dense layer) are trained by a Hebbian rule, so that ${\bf W}_{sd} = \sum_n {\bf s}_n^\top {\bf d}_n$ and  ${\bf W}_{ds} = \sum_n {\bf d}_n^\top{\bf s}_n$ the weights ${\bf W}_{da}$ are random and fixed, and ${\bf W}_{ad}$ are also held fixed (after being set as a pseudoinverse on the random ${\bf W}_{da}$, so these may also be viewed as random). Thus, the query weights are partly random, fixed, and coupled with the keys, and partly learned through Hebbian plasticity. The choice of modular attractor states and random projections (key weights) from attractor to dense layer pre-defines both well-separated keys and a NN operation for robust error-correction and matching of noisy queries with keys. Though learning is heteroassociative and values are stored separately from keys and queries, unlike Hopfield networks, the networks are content-addressable.  
\end{kbox}

In contrast to the adaptive addresses used in the Tolman-Eichenbaum Machine, Sharma, Chandra and colleagues \citep{sharma22,chandra25} proposed neurobiologically motivated models (MESH and Vector-HaSH) that use random projections of modular fixed point (attractor) networks to address memories (see Box \ref{box:hashmesh} for details). The fixed points within each modular attractor, which are fixed and not learned, may be sparse and random \citep{sharma22} or possess a 2-dimensional geometry \citep{chandra25} such as the hexagonally periodic activity patterns of grid cells. Consisent with the non-learned assumption, grid cells exhibit a fixed relational organization regardless of environment and behavioral state \citep{yoon13, gardner19, trettel19, gardner22}. The same relationships are re-used to encode the organization of non-spatial variables \citep{constantinescu16,aronov17,killian12}. The fixed modular attractor networks in the models randomly and densely project into a larger network (the hippocampus in Vector-HaSH) with self-consistent return projections, forming a large scaffold of quasi-random stable fixed point address states. Thus, we may view MESH and Vector-HaSH as fitting the diagram of Fig. 1 (right), except that the hidden layer in the diagram is replaced by a bipartite scaffold circuit (modular attractor network bidirectionally coupled to the densely connected layer) that computes nonlinear robust NN search to compute $\sigma(S({\bf K},{\bf Q}))$, with keys that are random and fixed. In Vector-HaSH, keys reside in hippocampus and the memory values are reconstructed at the sensory input layer (cortex). 

The advantages of this construction, beyond standard key-value constructions, are that (i) the address space is large, and (ii) the scaffold fixed points possess large and uniform basins of attraction. These properties allow many memories to be robustly addressed, with error correction and without the ``memory cliff' that afflicts many memory models combine keys and values \citep{amit87,mccloskey89} or learn the keys \citep{chandra25}: 
after a threshold number of inputs have been stored, retrieval performance crashes. In contrast, MESH and Vector-HaSH produce graceful degradation as multiple memories share overlapping addresses, much like human memory \citep{barnes59}. Remarkably, these models also outperform a flexible encoder trained to minimize reconstruction error \citep{radhakrishnan20}, suggesting that the entorhinal-hippocampal system may be a highly effective memory addressing system discovered by evolution.

The 2-dimensional and geometric aspect of Vector-HaSH over MESH becomes relevant for memorizing (possibly discrete) items embedded, remembered, or retrieved through traversal of a continuous low-dimensional space. This may involve sequential episodic memories (where time is the continuous dimension) or spatial memories (where space is the continuous dimension).

\section{Evidence from psychology and neuroscience}

In this section, we review several lines of evidence suggestive of key-value memory in the brain. Our review is organized around the following claims central to the theory:
\begin{enumerate}
    \item Memories are stored indelibly but subject to retrieval interference.
    \item Memories are addressed by representations (keys) that are distinct from the representations of memory content (values). The representational structure of the keys is optimized for discriminability, whereas the representational structure of the values is optimized for fidelity.
    \item The information stored in keys is not available to conscious access (i.e., recall). In other words, the brain uses keys to recall values but cannot recall the keys themselves.
\end{enumerate}

\subsection{Retrieval interference, not erasure, is the principal limiting factor in memory performance}
\label{sec:interference}

An important implication of key-value memory systems is that memory performance is constrained primarily by the ability to retrieve relevant information, not by storage capacity. In this section, we review some of the theoretical arguments and empirical evidence that this assumption can be plausibly applied to the brain.

A number of attempts have been made to estimate the storage capacity of the human brain \citep{dudai97}. Depending on different assumptions about numbers of neurons and connectivity, estimates can range from $10^{7}$ to $10^{15}$ bits. The total number of bits arriving from sensory systems with minimal compression are estimated to range from $10^{13}$ to $10^{17}$. These rough numbers suggest that with adequate compression, storage capacity may not be the strongest constraint on memory performance.

Perhaps more persuasively than these theoretical arguments, we can make the case based on observations about behavior. If memory storage has reached its capacity limit, then it is impossible to store new information without removing some old information. This implies that old information should become permanently inacessible at some point. In contrast, studies of human memory demonstrate that memories can be stored over decades, despite being rarely rehearsed \citep{bahrick75,bahrick91,conway91,maxcey21}.

One might try to explain these findings by arguing that memory storage has not reached the capacity limit, but then it would be very challenging to explain forgetting over the much shorter intervals studied in experiments on list memory. When presented with a list of random words, the proportion of recalled words declines with list length (typically in the range of 5 to 20 items). If this was due to removal of items from memory, then one would expect catastrophic forgetting over intervals of decades.
Furthermore, a series of experiments by Shiffrin \citep{shiffrin70} demonstrated that, when presented with a sequence of lists and then asked to recall the list before the most recently presented list, performance depended not on the length of the most recent list but only on the length of the list being recalled. This suggests that forgetting is not due to displacement of old items by items from the last list. The limiting factor is retrieval interference from other items on the same list.

Using word-location pairs, Berens and colleagues \citep{berens20} separately estimated memory accessibility (whether or not a location is recalled at all given a word cue) and precision (the variance of angular errors). They found that accessibility, but not precision, declined as a function of the retention interval. Similar results have been reported using memory for real-world events \citep{diamond20}. Thus, memories do not melt into oblivion, but rather disappear from view. When they come back into view, they are as sharp as they were before they disappeared.

Supposedly lost memories can be found when appropriate retrieval cues are used \citep{wagenaar86}, when the number of retrieval attempts increases \citep{buschke74,roediger78}, or even spontaneously after a sufficiently long delay \citep{payne87}. This holds not only for standard memory tasks in healthy subjects but also for retrograde amnesia induced experimentally or by neurological damage. Spontaneous ``shrinkage'' of amnesia is a common clinical observation following brain trauma \citep{kapur99}, presumably due to the restoration of memory access. In laboratory studies, amnesia can be induced experimentally in a range of ways, such as electroconvulsive shock, hypothermia, protein synthesis inhibition, and lesion or inactivation of the hippocampus, with recovery also induced in a range of ways \citep{lewis1968recovery,riccio84,miller21}. For example, protein synthesis inhibition following classical conditioning typically eliminates conditioned responding on long-term memory tests. However, delayed testing sometimes reveals recovery of performance \citep{flexner66,serota71,squire72}. It is even possible to restore performance using the amnestic agent itself \citep{bradley88,briggs13,gisquet15}.

These observations about amnesia mirror well-known phenomena in classical conditioning. Extinction of a previously conditioned stimulus (i.e., presenting the stimulus in the absence of the unconditioned stimulus) causes a decline in conditioned responding, eventually reaching baseline levels. This decline is transient: conditioned responding can return spontaneously \citep{pavlov27}, or can be induced by a single ``reminder'' of the unconditioned stimulus \citep{rescorla75}.

In summary, considerable evidence suggests that failures of remembering primarily arise from failures of retrieval, typically due to interference from other memories. Memories thought to be lost can later be found under the right conditions. We will explore this idea computationally when we discuss model simulations.

\subsection{Distinct representations of keys and values}

The influential ``Complementary Learning Systems'' framework holds that there is a division of labor between the hippocampus and neocortex \citep{mcclelland95,oreilly02}, with the hippocampus specialized for episodic memory (remembering events that occurred in a specific spatiotemporal context) and a set of neocortical areas (sometimes referred to as ``association cortex'') that are specialized for semantic memory (remembering regularities that generalize across episodes). This framework has also influenced the design of artificial intelligence systems \citep{kumaran16}. In this section, we will argue that this framework can be understood in terms of key-value memory.

Rather than thinking about the hippocampus as directly storing memory content, we can alternatively conceptualize it as storing keys and matching queries to keys, which address memory content stored in neocortex. This view emphasizes the point that episodic memories need to be bound to semantic content---otherwise, they're essentially empty vessels, as in cases of semantic demantia, where degeneration of anterior temporal lobe and prefrontal areas produces profound semantic impairments despite relatively intact recognition memory for recent experiences \citep{graham99}.

If the hippocampus provides the keys necessary for activating neocortical values, then we would expect a causal interplay between the two. Indeed, there is empirical evidence for the following facts: (i) cortical encoding-related activity is reinstated at the time of memory retrieval; (ii) cortical reinstatement depends on the hippocampus; and (iii) the reinstatement is necessary for memory retrieval \citep{staresina12,bosch14,tanaka14,danker17,pacheco19,hebscher21}.

Another line of evidence comes from studies of generalization. In the absence of the hippocampus, neocortical values cannot be accessed in a targeted way, leading to overgeneralization. A study by Winocur and colleagues \citep{winocur09} offers a good example. When trained to anticipate a shock in Context A, rats specifically freeze in Context A but not in Context B when tested a day later. When tested a week later, rats show a generalization effect (loss of context specificity), freezing in both contexts. This generalization might arise because of interference from memories acquired during the intervening time. Context specificity can be restored by ``reminding'' the rat of the original memory (briefly placing it back in Context A). This reminder effect can be interpreted as activation of the appropriate address given a highly specific cue. Importantly, the reminder effect disappears in hippocampal-lesioned rats, consistent with the claim that the hippocampus stores the keys necessary for targeting specific memories (see also \citep{wiltgen10} for converging evidence).

The idea that the hippocampus stores keys was to a large extent anticipated by the hippocampal memory indexing theory \citep{teyler86}, which proposed that the hippocampus indexes memory content stored in neocortical areas. Since it was first proposed, new techniques have uncovered much more detailed support for the theory \citep{teyler07,goode20}. In particular, the advent of activity-dependent labeling and optogenetic manipulation have enabled the identification of ``engram cells'' in the hippocampus which are causally linked to specific memories \citep{liu12,ramirez13}. Goode and colleagues \citep{goode20} interpret hippocampal engrams as indices (in the sense of Teyler and DiScenna), linking together information stored in a distributed network of neocortical areas.

Several brain-wide engram mapping studies in rodents have reported a collection of neocortical (as well as subcortical) areas that are active during both encoding and retrieval, and therefore qualify as engrams \citep{vetere17,roy22}. What makes the hippocampus special is its role as a hub region with high connectivity to neocortical regions \citep{battaglia11}. This allows the hippocampus to exert strong control over neocortical regions. In addition, hippocampal engrams are highly sparse (involving a small subset of hippocampal cells) and conjunctively tuned to multiple sensory and cognitive inputs; these features make hippocampal engrams well-suited to encoding episodically unique memories. A recent study of food-caching birds \citep{chettih24} provides a particularly dramatic demonstration: hippocampal ensembles encoded unique representations for over 100 cache sites (including multiple caches at the same location), which were reactivated upon memory retrieval.

If hippocampal representations are optimized for discriminating between distinct episodes in a cue-dependent manner, then we should expect changes in these representations under different retrieval demands. For example, overlapping routes in spatial navigation need to be discriminated in order to avoid confusion. Chanales and colleagues \citep{chanales17} showed that this situation produces repulsion of hippocampal representations specifically for overlapping locations (see also \citep{wanjia21}). The repulsion effect emerges gradually over the course of learning, ultimately reversing the objective similarity relations between locations. The strength of the repulsion effect also correlated with behaviorally measured discrimination accuracy.

Note that optimization for discriminability is only part of the story, since the hippocampus receives noisy inputs which may activate the wrong keys. The hippocampus needs to do some error correction / pattern completion in order to ``clean up'' the retrieved keys. This may be accomplished by attractor dynamics such that nearby inputs get mapped to a corresponding attractor (pattern completion), while more distant inputs are separated into distinct attractors (pattern separation). It remains to be seen if such attractor dynamics are implemented by the recurrent hippocampal-entorhinal loop \citep{chandra25} (Box 1) or by the proposal that both occur within hippocampus, with pattern separation by dentate gyrus and pattern completion by area CA3 \citep{rolls13}.

In summary, evidence suggests a division of labor between key encoding in the hippocampus and value encoding in the neocortex. Hippocampal keys are optimized for discrimination, whereas neocortical values are optimized for encoding semantic regularities. 

\subsection{Values, but not keys, are available for recall}

Keys store information which is never overtly recalled. Testing this hypothesis is challenging because it is possible that some information stored in keys is also stored in values. Clear support comes from evidence that there is information used to guide retrieval (putatively stored in keys) which nonetheless is not available for overt recall. More technically, we characterize ``overt recall'' of a key as the ability to report some aspect of the key vector. The evidence presented below suggests that keys can be matched to queries without making the content of the keys available for report.

Many of us are familiar with the experience of having a memory at the ``tip of the tongue''---stored in memory yet temporarily unrecallable \citep{brown66,brown91}. Closely related is the ``feeling of knowing''---a subjective judgment about the likelihood of subsequently recognizing items which are presently unrecallable. The typical study procedure is to present subjects with difficult general knowledge questions, elicit feelings of knowing for questions they cannot presently answer, and then subsequently test their ability to recognize correct answers to the questions. An important finding from this literature is that feelings of knowing predict (though not perfectly) subsequent recognition \citep{hart65,freedman66}, indicating that people are able to judge whether some piece of information is stored in memory despite not being able to retrieve it. Similar results have been found with cued recall tests \citep{gruneberg74}. Another clue comes from studies examining response times: Reder \citep{reder87} found that people could report answerability for trivia questions faster than they could report the answers, again indicating that metamemory does not require retrieval of memory content. Furthermore, feelings of knowing (but not recall) can be spuriously increased by increasing cue familiarity \citep{reder92,schwartz92}, possibly due to increased key-query match without increased probability that the correct keys are matched.

These phenomena are broadly consistent with the hypothesis that key-query matching can be used to judge whether some information (e.g., the answer to a question) is stored in memory, without accessing the memory content (value) itself. This idea has appeared in the human long-term memory literature under various guises; for example, Koriat \citep{koriat77} discuss the tip-of-the-tongue phenomenon in terms of ``pointers'' (cues that specify a memory address without specifying memory content), whereas Morton and colleagues \citep{morton85} use ``headings'' to refer to essentially the same idea.

A pointer system has also been invoked to understand short-term memory: rather than storing transient copies of long-term memories, short-term memory might store pointers which refer to information in long-term memory \citep{ruchkin03,norris17}. This may explain why people can detect changes even when they cannot report what exactly has changed \citep{ball15}, analogous to the ``butcher on the bus'' phenomenon in long term memory \citep{mandler80}, where people can recognize familiar items without being able to recollect any details about them. Both change detection without identification and familiarity without recollection might be accomplished using keys (pointers) that are content addressable but do not obligatorily activate their associated values \citep{chandra25}.

Many standard models of memory cannot check whether information is stored without retrieving that information at least partially. This is because most models base recognition memory judgments on the match between the cue and stored content; there is no separate representation of keys that can be used to check whether something is stored in memory. This makes it challenging for these models to explain why people can be knowledgeable about what is stored in memory without recalling it.

\section{Illustrative simulations}
\label{sec:simulations}

In this section, we provide two simulations that illustrate some of the distinctive characteristics of key-value models highlighted above.
Further implementation details can be found in our public code repository, available at \url{https://github.com/kazuki-irie/kv-memory-brain}.

\begin{figure}
    \begin{center}
    \includegraphics[width=1\textwidth, scale=1]{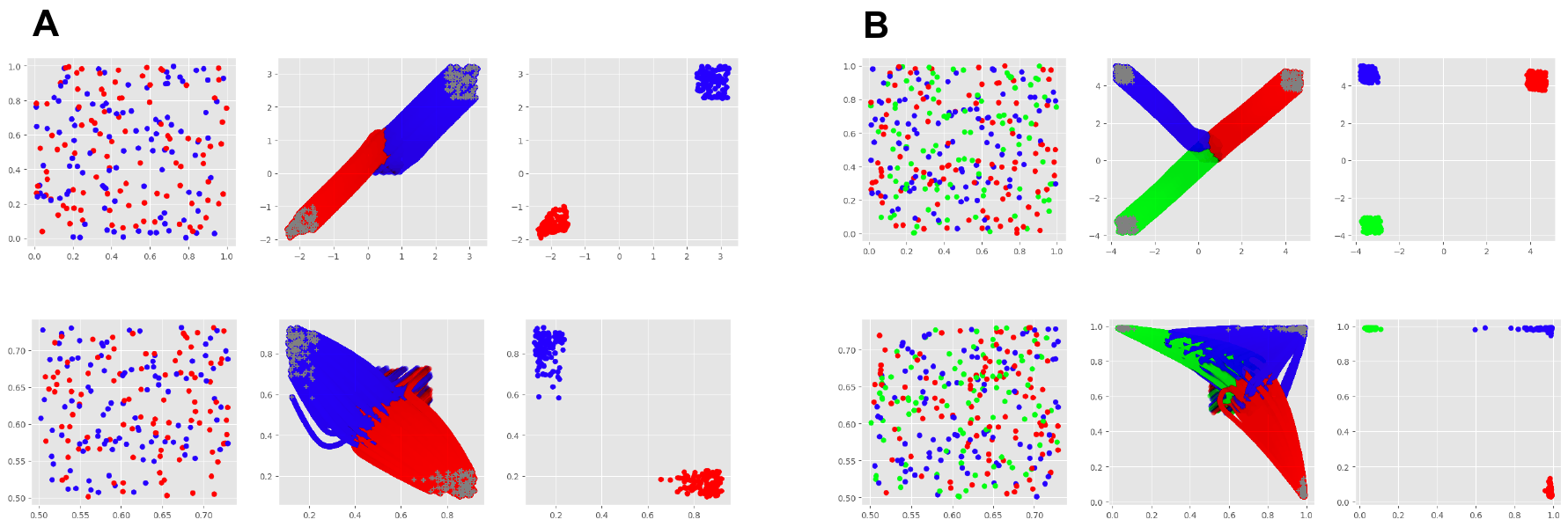}
    \end{center}
    \caption{\textbf{Optimization of key and value representations}. Each point represents an event in the memory and belongs to one of (A) two or (B) three classes, represented by different colors. In each case, the evolution of key (Top row) and value (Bottom row) representations during the optimization process is shown; each row shows (Left) Random initialization, (Middle) trajectory of representations during the optimization process, with the final positions marked by gray points, (Right) final configuration. We observe that keys are optimized for retrieval/separability, while values are optimized to store the memory content.}
    \label{fig:representations}
\end{figure}

\subsection{Distinct representations for keys and values}
\label{sec:simulation_key_value}

As we discussed earlier, one of the essential properties of key-value memory is the separate representations allocated for keys and values, which can be optimized for their specific roles in retrieval and storage, respectively.
Here we present a toy simulation that illustrates this property.

We consider a minimal key-value model whose trainable parameters are a set of pairs of key and value vectors (i.e., in this model, we skip the step of mapping inputs to keys and values, and directly examine the properties of key/value representations).
We set both keys and values to be 2-dimensional (2D) vectors which can be easily visualized in the 2D space.
The model can take an arbitrary 2D vector as an input which is treated as a query; the query is compared to all the keys through dot product to obtain a similarity score for each key (as in Eq.~\ref{eq:scaled_dot_prod}). The resulting scores are normalized by applying the softmax function (Eq.~\ref{eq:softmax}) to obtain the final ``attention'' scores.
The output of the model is the weighted average of value vectors using these attention scores (Eq.~\ref{eq:retrieval}).

To train the model, we randomly assign each key/value pair to a class; we test two settings using either two or three classes in total.
Each class has a fixed feature vector (representing some specific object or event): in the two-class case,
the feature vectors for Class `0' and `1' are vectors $(0, 1)$ and $(1, 0)$ in the 2D space, respectively;
in the three-class case, we additionally have a third class, Class `2' with $(1, 1)$ as its feature vector.
The task of the model is to output the correct class feature vector when one of the keys is fed to the model as input. We apply the sigmoid function to the value vectors to ensure their effective values are between 0 and 1.
The key and value vectors are initialized with a uniform distribution between 0 and 1, and our goal is to examine what key and value representations emerge when this key-value model is optimized for this simple retrieval task. 
We use the mean squared error loss and the gradient descent algorithm, as is commonly used in modern deep learning.

The results are shown in Figure \ref{fig:representations}.
We observe that key and value representations effectively exhibit different optimal configurations.
The key configuration is optimized for softmax-based discrimination, facilitating effective retrieval: as we can see in Figure \ref{fig:representations}A, Top row (the two-class case), the two classes take two opposite quadrants (which are highly distinguishable through dot product and softmax). The trend is similar for the three-class case (Figure \ref{fig:representations}B, Top row).
In contrast, the value representations are optimized to represent the class features (i.e., the memory content). For example, in the two-class case (Figure \ref{fig:representations}A, Bottom row), they are optimized to represent $(0, 1)$ for Class `blue' and $(1, 0)$ for Class `red'. Again, the trend is similar for the three-class case (Figure \ref{fig:representations}B, Bottom row).
This illustrates how the key-value memory architecture allows for separate representations for keys and values, optimized for retrieval and storage, respectively.

\subsection{Forgetting as retrieval failure, and recovery by memory reactivation}
\label{sec:simulation_forgetting}

Another distinctive property of the key-value memory we highlighted earlier is that, as recall relies on successful retrieval, forgetting can be conceptualized as retrieval failure; that is, even when a memory trace (i.e., a key/value pair corresponding to an event) is stored in memory, forgetting can still occur when retrieval fails due to interference/inaccessibility.
This also implies that, if we manage to fix the failure in the retrieval process, there is a hope for recovering the corresponding memory without having to relive the event itself.
Here we present a simulation illustrating these phenomena using an artificial neural network that learns two tasks sequentially (the so-called continual learning scenario), and examine it through the lens of key-value memory.
We also highlight how this simple experiment echoes neurobiological findings on optogenetic recovery of memory following retrograde amnesia \citep{ryan2015engram,roy2017silent}.

We train a simple feedforward neural network on two binary image classification tasks sequentially.
Using two classic image datasets commonly used in deep learning, MNIST and FashionMNIST \citep{lecun1998mnist,xiao2017fashion}, we construct the two toy tasks that involve classifying images from the two first classes of MNIST (digit `0' vs.~`1') and FashionMNIST (`T-shirt' vs.~`Trouser'), respectively.
The model has one hidden layer (i.e., it has two linear transformations: input-to-hidden, and hidden-to-output mappings with a rectifier linear activation function in-between). The input images are grayscale and their dimension is 28$\times$28; they are flattened to yield a 784-dimensional vector accepted by the input layer. We set the hidden-layer dimension to 64, and the model output size is 4 (for two times two-way classification tasks). The model is trained on the cross-entropy loss using the gradient descent algorithm; by applying the dual formulation described earlier, each linear layer in the model can be formalized as a key-value system keeping memory traces of the entire learning experience by storing layer inputs as keys and error signals as values for every learning step.

In this sequential learning setting, we first train the model on Task 1 (MNIST) for 5 epochs, after which the model is trained for another 5 epochs on the training dataset of Task 2 (FashionMNIST) without having access to the Task 1 training data anymore.
While such a training process produces a set of weight matrices for neural networks in their conventional form, in the key-value memory view, the final model consists of a sequence of key/value vectors; in this specific scenario of two-task sequential learning, each key/value pair belongs to either Task 1 (MNIST) or Task 2 (FashionMNIST) learning experiences.

\begin{figure}
    \begin{center}
    \includegraphics[width=1\textwidth, scale=1]{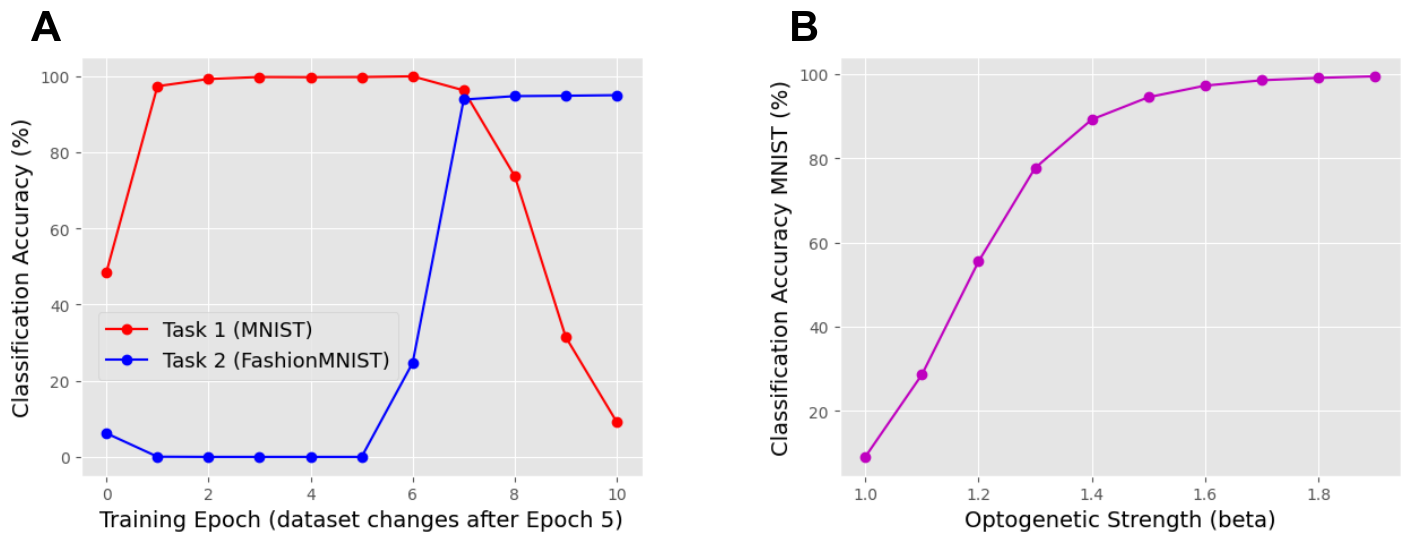}
    \end{center}
    \caption{\textbf{Forgetting and reactivation of memory events}. A one-layer feedforward neural network is trained on two tasks sequentially, Task 1 and 2, constructed using the MNIST and FashionMNIST datasets, respectively. (A) The evolution of the test classification accuracy for the two tasks as a function of training epochs. After epoch 5, the training dataset changes from Task 1 to Task 2; resulting in forgetting of Task 1 as the model learns Task 2. (B) The accuracy of the trained model on Task 1 as a function of the value of the artificial scaler $\beta$ used to amplify the keys in all key-value memory pairs corresponding to Task 1 learning.}
    \label{fig:amnesia}
\end{figure}

Figure \ref{fig:amnesia}A shows the evolution of model performance on the test set of the two tasks as a function of training epochs.
We observe that the model achieves $\sim$99\% test accuracy on Task 1 (MNIST) in the first 5 epochs corresponding to Task 1 learning, but this performance drops to $\sim$9\% after learning Task 2 (FashionMNIST) until the model achieves $\sim$95\% test accuracy on Task 2 (we deliberately train the model long enough to observe forgetting).
This amnesic phenomenon, reminiscent of catastrophic forgetting in neural networks \citep{mccloskey89,ratcliff1990connectionist,french1999catastrophic}, is intriguing given the key-value memory view of the model, in which the key/value memory pairs belonging to Task 1 remain part of the model parameters, explicitly and indefinitely. By using the same terminology as in neuroscience experiments \citep{ryan2015engram}, these Task 1-related key/value memories became ``silent'' after Task 2 learning.

An interesting question is whether we can reactivate these 
silent key/value memories to recover the model's performance on Task 1 without any retraining on Task 1, akin to how experimental neuroscientists successfully reactivated ``silent engrams'' through an optogenetic procedure \citep{roy2017silent}.
For this, we introduce a single positive scalar $\beta \geq 1$ (an ``optogenetic strength'') to multiply/amplify the keys (or, equivalently in this model, the values) in all key-value pairs corresponding to Task 1 in the trained model.
Figure \ref{fig:amnesia}B shows the performance of the model on Task 1 (MNIST) as a function of the optogenetic strength $\beta$ ($\beta=1$ corresponds to no intervention on the model).
We observe that by simply increasing $\beta$ (i.e., by artificially ``reactivating'' the existing key-value memories corresponding to the model's prior MNIST learning experiences), the model becomes capable of solving MNIST again without any retraining.

This simulation not only illustrates the core property of the key-value memory, where forgetting can be attributed to retrieval failure, but also resonates with the neuroscientific findings of silent memory engrams in retrograde amnesia and their recovery through artificial reactivation.

\section{Conclusions}

Our goal in this paper was to connect ideas about key-value memory across artificial intelligence, cognitive science, and neuroscience. We have argued that the brain might plausibly implement a key-value memory system in the division of labor between hippocampus and neocortex. We have also argued that a number of behavioral findings from memory studies (e.g., the ability to report item familiarity without recollection) is consistent with a key-value architecture.

Presently, the connections we have highlighted are speculative. On the empirical side, we hope that more direct experimental tests will be undertaken. For example, the key-value architecture implies that repulsion effects in long-term memory should be reversible---a prediction that has not yet been tested. On the theoretical side, we hope that more biologically detailed models will be developed which can explain the plethora of findings discussed in previous sections.

Our paper was motivated by the fact that key-value memory seems to be an important ingredient in the success of several modern artificial intelligence systems such as transformers and fast weight programmers. The idea that the brain may implement something like this indicates evidence for convergence \citep{gershman24} and suggests that this is a promising direction for exploring brain-like mechanisms that can power intelligent systems.

\subsection*{Acknowledgments}

The authors are grateful for support from the Kempner Institute for the Study of Natural and Artificial Intelligence, and from the Department of Defense MURI program under ARO grant W911NF-23-1-0277.

\subsection*{Declaration of interests}

The authors declare no competing interests.

\bibliographystyle{unsrtnat}
\bibliography{bib}

\clearpage
\section*{STAR $\bigstar$ METHODS}

\subsection*{KEY RESOURCES TABLE}
\begin{table*}[h]
\footnotesize
\begin{center}
\begin{tabular}{lll}
\toprule
 RESOURCE & SOURCE & IDENTIFIER \\ \midrule \midrule
Data & &  \\  \midrule
MNIST & LeCun et al.~\citep{lecun1998mnist} & \url{https://ossci-datasets.s3.amazonaws.com/mnist} \\
Fashion MNIST & Xiao et al.~\citep{xiao2017fashion} & \url{http://fashion-mnist.s3-website.eu-central-1.amazonaws.com} \\ \midrule \midrule
Software  & & \\  \midrule
PyTorch 2.5.1 & Paszke et al.~\citep{Paszke19} & \url{https://pytorch.org/} \\
Code for this paper & This paper & \url{https://doi.org/10.5281/zenodo.14920916} \\
\bottomrule
\end{tabular}
\end{center}
\end{table*}

\subsection*{RESOURCE AVAILABILITY}

\paragraph{Lead contact}
Further information and requests for resources should be directed to the lead contact, Samuel J. Gershman (gershman@fas.harvard.edu).

\paragraph{Materials availability}
This study did not generate new materials.

\paragraph{Data and code availability}
Our code is publicly available online at:  \url{https://github.com/kazuki-irie/kv-memory-brain}.
This repository contains the code required to reproduce all the simulation results presented in this paper, including Figures \ref{fig:representations} and \ref{fig:amnesia}.

\subsection*{METHOD DETAILS}

Our code was implemented using PyTorch \citep{Paszke19}.

\paragraph{Distinct representations for keys and values} 

In both the two-class and three-class settings, 100 two-dimensional vectors were uniformly sampled from the range [0, 1] for each key and value within each class, resulting in a total of 200 key/value pairs for the two-class case (Figure \ref{fig:representations}A) and 300 for the three-class case (Figure \ref{fig:representations}B).

The optimization was performed using the Adam optimizer \citep{kingma15} with a learning rate of 3e-4 for 5,000 steps in the two-class setting and 10,000 steps in the three-class setting.

The experiments were conducted using the free version of Google Colab with a CPU.
The python notebooks to reproduce the results are \url{https://github.com/kazuki-irie/kv-memory-brain/blob/master/2classes_key_value_optimization.ipynb} and \url{https://github.com/kazuki-irie/kv-memory-brain/blob/master/3classes_key_value_optimization.ipynb} for the two-class and three-class cases, respectively.

\paragraph{Forgetting as retrieval failure, and recovery by memory reactivation}

Both the MNIST and Fashion MNIST datasets were used without modifications. We used the class `0' and `1' images from each dataset; resulting in 12,665 training images for MNIST and 12,000 for Fashion MNIST. Note that MNIST does not have an equal number of examples per class: 5,923 for class `0' and 6,742 for class `1.' We left this imbalance as is, as it is irrelevant to the main goal of our experiment.
The corresponding test sets consist of 2,115 and 2,000 images for MNIST and Fashion MNIST, respectively.

The initial weights of the two linear layers within the feedforward neural network were uniformly sampled from a uniform distribution over the range [-$a$, $a$] where $a = 1/\sqrt d_\text{in}$, and $d_\text{in}$ denotes the input dimension of the corresponding layer. No bias was applied in either of the two layers.
The optimization was conducted using the standard stochastic gradient descent algorithm using a learning rate of 6e-4 and a batch size of 128 images.
We refer to one ``epoch'' of optimization as a single iteration over the entire training dataset.

The experiments were conducted using the free version of Google Colab with a T4 GPU.
The python notebook to reproduce the corresponding results can be found at: \url{https://github.com/kazuki-irie/kv-memory-brain/blob/master/Forgetting_and_recovery.ipynb}.

\end{document}